\documentclass[twocolumn]{aastex6}
\usepackage{amsmath,amstext}
\usepackage{amssymb}
\usepackage{epsfig}
\usepackage{color}
\usepackage{hyperref}
\usepackage{apjfonts}
\usepackage[flushleft]{threeparttable}
\usepackage{float}

\def\cha    {{\em Chandra}\/}

\def\xmm        {{\em XMM-Newton}\/}

\begin{document}

\title{Constraining the magnetic field in the TeV halo of Geminga with X-ray Observation}
\author{Ruo-Yu Liu\altaffilmark{1}, Chong Ge\altaffilmark{2,4}, Xiao-Na Sun\altaffilmark{3}, Xiang-Yu Wang\altaffilmark{3}}
\altaffiltext{1}{Deutsches Elektronen Synchrotron (DESY), Platanenallee 6, D-15738 Zeuthen, Germany}
\altaffiltext{2}{Purple Mountain Observatory, Chinese Academy of Sciences, Nanjing 210008, China}
\altaffiltext{3}{School of Astronomy and Space Science, Nanjing University, Nanjing 210093, China}
\altaffiltext{4}{Department of Physics and Astronomy, University of Alabama in Huntsville, Huntsville, AL 35899, USA}

\begin{abstract}
Recently, the High Altitude Water Cherenkov (HAWC) collaboration reported the discovery of the TeV halo around the Geminga pulsar. The TeV emission is believed to originate from inverse Compton scattering of pulsar-injected electrons/positrons off cosmic microwave background photons. In the mean time, these electrons should inevitably radiate X-ray photons via the synchrotron radiation, providing a useful constraint on the magnetic field in the TeV halo. In this work, we analyse the data of \xmm\ and \cha, and obtain an upper limit for the diffuse X-ray flux in a region of $600''$ around the Geminga pulsar, which is at a level of $\lesssim 10^{-14}\rm erg\,cm^{-2}s^{-1}$. Through a numerical modelling on both the X-ray and the TeV observations assuming isotropic diffusion of injected electrons/positrons, we find the magnetic field inside the TeV halo is required to be $<1\mu$G, which is significantly weaker than the typical magnetic field in the interstellar medium. The weak magnetic field together with the small diffusion coefficient inferred from HAWC's observation implies that the Bohm limit of particle diffusion may probably have been achieved in the TeV halo. We also discuss alternative possibilities for the weak X-ray emission, such as the hadronic origin of the TeV emission or a specific magnetic field topology, in which a weak magnetic field and a very small diffusion coefficient might be avoided.
\end{abstract}

\section{Introduction}
Recently, the HAWC collaboration reported the discovery of TeV gamma-ray haloes around about $10^\circ$ of two nearby pulsars, i.e, Geminga and Monogem\citep{HAWC17_Geminga}. Given that pulsars are promising accelerators of cosmic ray electrons/positrons (hereafter, for simplicity we do not distinguish positrons from electrons unless specified), the extended TeV emissions may probably arise from inverse Compton (IC) scatterings of $\sim 10-100$\,TeV electrons off cosmic microwave background photons. By jointly modelling the surface brightness profiles (SBPs) of these two TeV haloes, the HAWC collaboration suggests a low diffusion coefficient of $D = (4.5 \pm 1.2) \times 10^{27}\,\rm cm^2/s$ at 100\,TeV in the TeV haloes.

The derived diffusion coefficient is more than two orders of magnitude smaller than the standard diffusion coefficient in the interstellar medium (ISM) that is inferred from the measurement of secondary-to-primary ratio in CR spectrum (i.e., the Baron-to-Carbon ratio, \citealt{Aguilar16}). \citet{Hooper17} pointed out that the measured electron spectrum by the H.E.S.S. extending to about 20 TeV disfavors a small diffusion coefficient throughout the bulk of the local interstellar medium. Then, a plausible scenario is the appearance of two diffusion zones between the pulsar and Earth, with the pulsar locating inside an inefficient diffusion zone of a size of a few tens parsecs, while the diffusion coefficient in the bulk of ISM is the standard one as inferred from CR measurement. Such a scenario has been proposed in some previous literature \citep{Hooper17b, Fang18, Profumo18, Tang18}. However, the cause of the inefficient diffusion zone is still not clear yet. The inefficient diffusion zone may be the parent supernova remnant of the pulsar and the turbulence is driven by the shock \citep{Fang18}, or it may be the relic pulsar wind nebula where the magnetic field is
higher than that in ISM and the magnetic topology could be complicated \citep{Tang18}. Alternatively, the low diffusion coefficient could be caused by the instability due to gradient of CRs injected from the pulsar \citep{Quenby18, Evoli18}. 

\begin{figure*}[htbp]
\begin{center}
\includegraphics[width=0.39\textwidth]{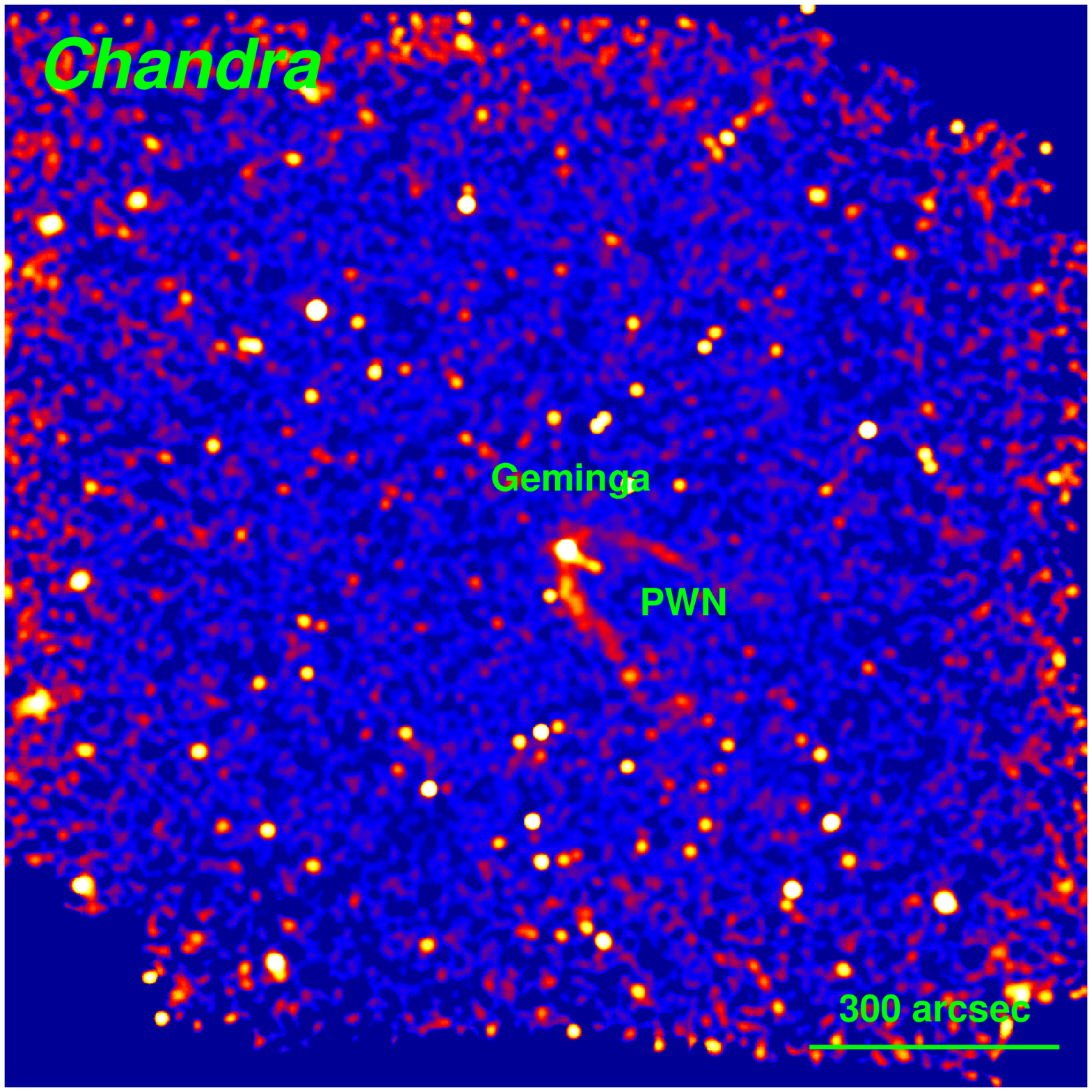}\quad \qquad 
\includegraphics[width=0.39\textwidth]{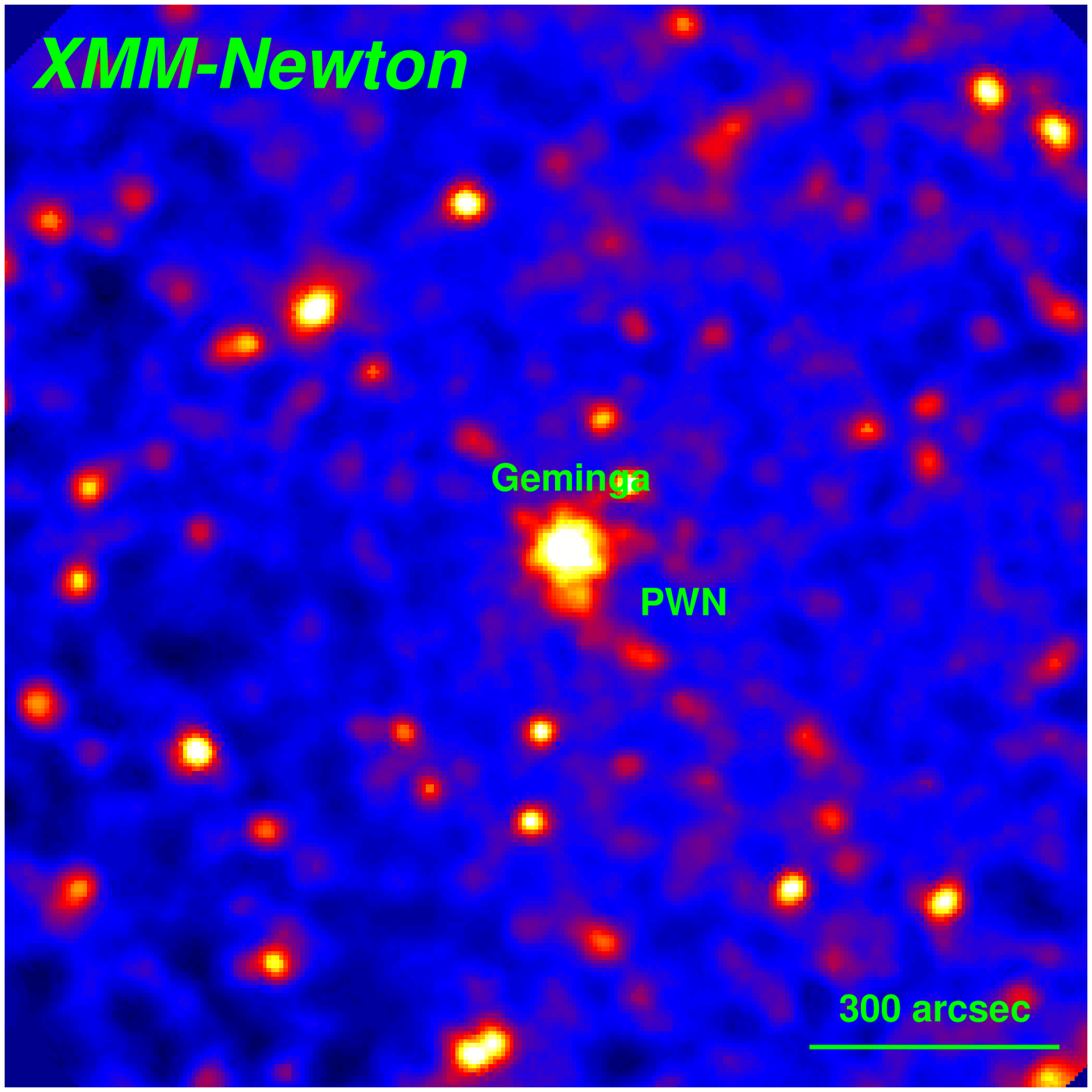}\\
\end{center}
\vspace{10pt}
\quad \quad
\includegraphics[width=0.43\textwidth]{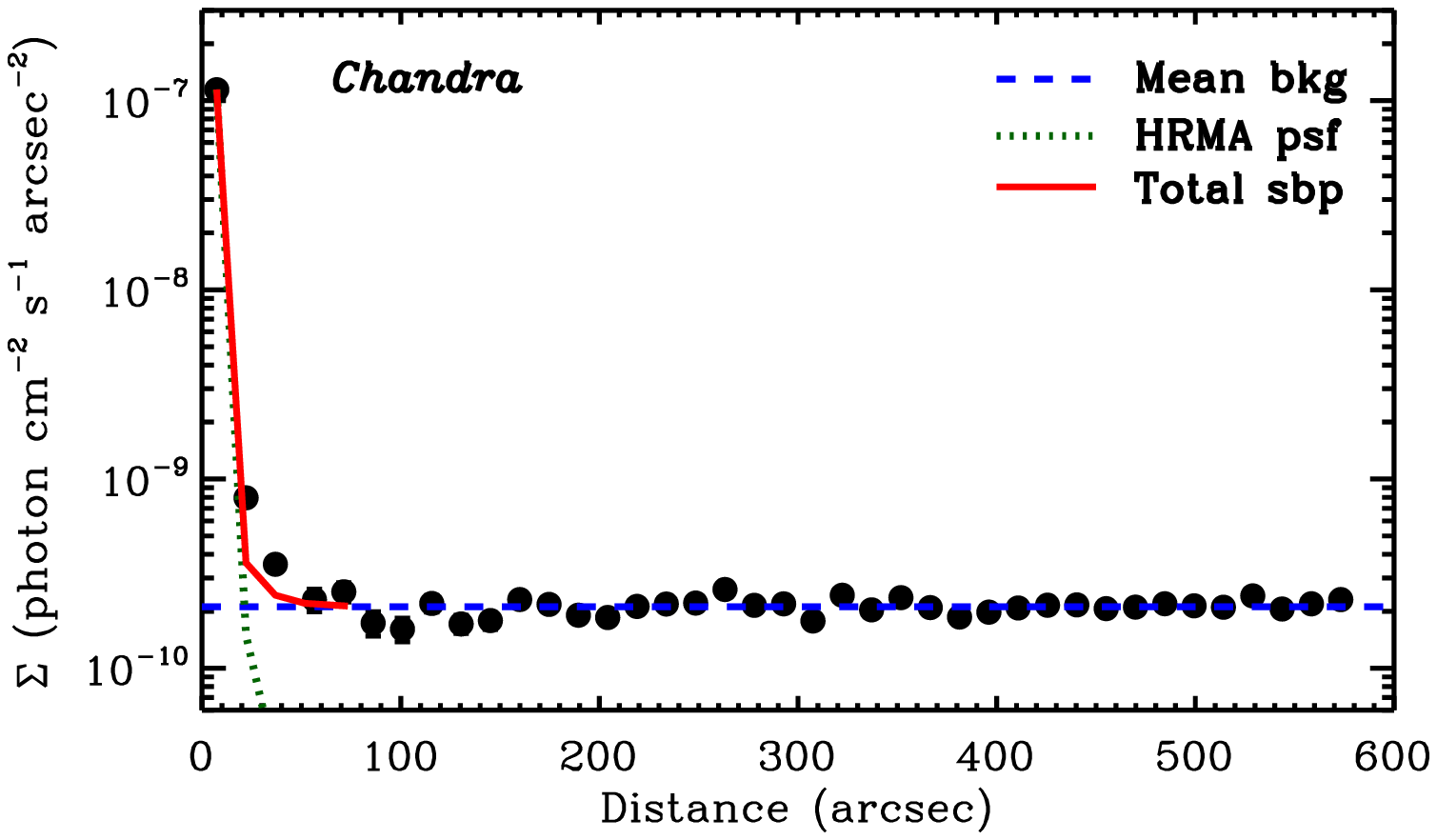}
\includegraphics[width=0.43\textwidth]{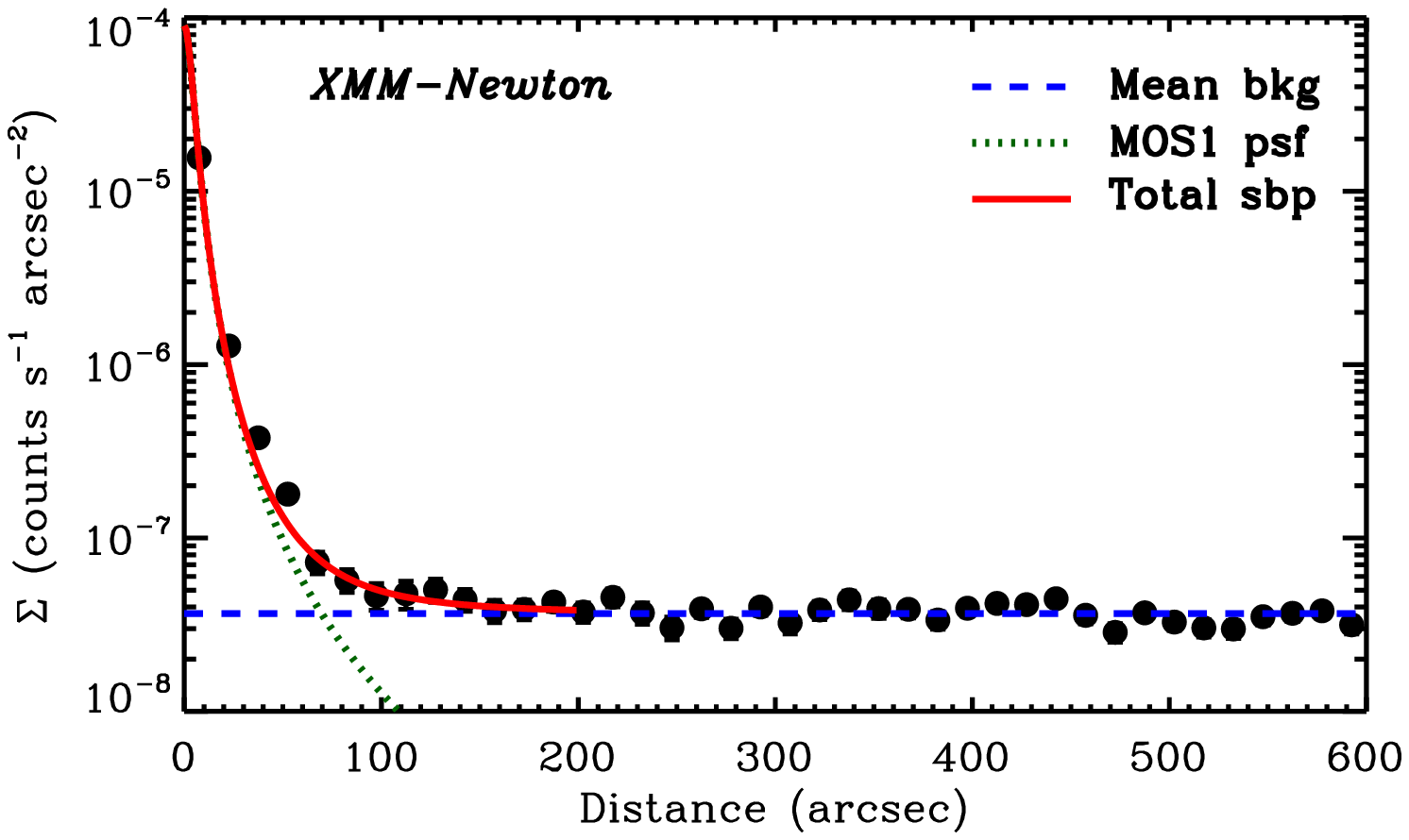}
	\caption{{\bf Upper panel}:\cha\ 0.7-2.0 keV and \xmm\ 0.7-1.3 keV instrumental background subtracted, exposure corrected images. The central bright Geminga and its PWN are labeled. {\bf Lower panel}: The diffuse SBP around Geminga. The central region of SBP is dominated by the instrumental PSF, while the outer region of SBP is flat. Thus it is dominated by the cosmic background shown as the blue dashed line.}
	 \label{fig:xray}
\end{figure*}

For any turbulence-driven mechanism, the key is the growth of the plasma instability. The derived diffusion coefficient by the HAWC collaboration around the pulsars implies the turbulence is nearly at saturation, with the perturbed magnetic field $\delta B$ comparable to the mean field $B$. Thus, the magnetic field in the TeV halo is very crucial to understand the origin of the low diffusion coefficient. On the other hand, we note that the energy of electrons that radiate $\sim 20$\,TeV photons via IC scattering off CMB photons is about 100\,TeV, i.e., $\epsilon_{\rm IC}\sim 20(E_e/100{\rm TeV})^2$\,TeV. These electrons will inevitably radiate in the magnetic field via synchrotron radiation and most likely give rise to a diffuse X-ray halo with a typical energy $\epsilon_{\rm syn}\sim 0.6(E_e/100{\rm TeV})^2 (B/3\mu \rm G)\,$keV with $E_e$ being the electron energy. The ratio between the diffuse X-ray flux and the diffuse multi-TeV flux from the same region is approximately equal to the ratio between magnetic field energy density in the TeV halo and the CMB energy density $U_{\rm CMB}$, i.e., $F_{\rm keV}/F_{10 \rm TeV}\simeq B^2/8\pi U_{\rm CMB}$. Thus, the diffuse X-ray flux provides a clue to understand the magnetic field in the TeV halo.

In this work, we will study the magnetic field via the X-ray emission around Geminga. The remainder of this paper is organized as follows: in Section~2, we present our analysis of the X-ray data on the region around Geminga. In Section~3, we obtain the upper limit of the magnetic field in the TeV halo based on the theoretical modelling of the TeV and the X-ray emission. In Section~4, we discuss the implications of our result and examine alternative interpretations. We give the conclusion in Section~5.

\section{Analysis of X-ray data}
\label{s:obs}
Table~\ref{t:obs} lists the \cha\ and \xmm\ observations used here. We reduce the X-ray data, following similar procedures in \citet[e.g.][]{Ge18, Ge18b} and briefly summarizing here, to study the diffuse X-ray emission around Geminga.

\subsection{Chandra}
We reduce the \cha\ ACIS-I data with the \cha\ Interactive Analysis of Observation (CIAO, version 4.9) and calibration database (CALDB, version 4.7.3). For each observation, we use the {\tt chandra\_repro} script with
VFAINT mode correction to reproduce a new level=2 event file. Then we use {\tt deflare} to filter the flares that deviating more than $3\sigma$ from the mean count rate. The exposures of clean and original time are also included in Table~\ref{t:obs}. Point sources are detected by {\tt wavdetect}. The point spread function (PSF) of \cha\ High-Resolution Mirror Assembly (HRMA) shown in Fig.~\ref{fig:xray} is modeled with the Chandra Ray Tracer (ChaRT) and {\tt simulate\_psf}. The instrumental stowed background 
are reprojected to match each of the ACIS-I chips, with 1) VFAINT cleaning; 2) point sources region masked; and 3) rescaled to the count rate in the
9.5-12 keV band. We use the {\tt merge\_obs} to produce the merged event and exposure maps from multiple observations. Then the merged event image is subtracted with the merged stowed background image, and further divided by the exposure map to get the flux image, from where we extract a SBP of the diffuse X-ray emission in 0.7-2.0 keV. Both the flux image and SBP are shown in Fig.~\ref{fig:xray}.

\subsection{XMM-Newton}
We process the \xmm\ MOS data with the Extended Source Analysis Software (ESAS; \citealt{Kuntz08, Snowden08}), as integrated into the \xmm\ Science Analysis System (version 15.0.0) with the associated Current Calibration Files (CCF). The positive-negative (pn) data is in the small-window operating mode with limited field of view (FOV), thus we do not analysis the pn data.
We reduce the raw event files from MOS with {\tt emchain} task. We use {\tt mos-filter} to filter out the flares from solar soft proton. The MOS CCDs that are damaged or in the anomalous state are excluded in downstream processing. We apply the point source positions from \cha\ with the \xmm\ PSF from {\tt calview} and \citet{Ghizzardi02}, as well as the additional point sources detected by {\tt cheese} in the outer radius, which is not covered by the \cha\ observations. We use {\tt mos-spectra} to produce event images and exposure maps. The instrumental background images are modeled with  {\tt mos$\_$back}. The residual soft proton background images are modelled with {\tt proton}. 
We combine the event images, background images, and exposure maps from multiple observations as well as from MOS1 and MOS2 with {\tt merge\_comp\_xmm}. Then we produce a flux image and extract a SBP in 0.7-1.3 keV.

\subsection{The diffuse X-ray emission}
The upper panel of Fig.~\ref{fig:xray} shows the X-ray emission from \cha\ and \xmm.
We then produce the SBPs after masking the point source (except the Geminga pulsar at the center) and the diffuse pulsar wind nebula (PWN) around Geminga (e.g., \citealt{Caraveo03, Posselt17}).
The diffuse X-ray SBP is shown in the lower panel of Fig.~\ref{fig:xray}. Excluding the central region affected by the instrument PSF, we do not find a flux drop toward larger radius as that in TeV band as measured by HAWC. Instead, the SBP is flat. We also compare the diffuse X-ray background within 10$^\circ$ around Geminga using the RASS R45 ($0.47-1.21$\,keV) flux. The fluxes from different radii are compatible with each other. Thus the contribution of  diffuse synchrotron radiation is not significant and the diffuse X-ray emission around Geminga is dominated by the cosmic background. We then estimate a $3\sigma$ upper limit flux for the diffuse synchrotron radiation, assuming a power-law model, with the index $\Gamma=2.0$. The choice of the power-law index does not affect our final estimate significantly given the narrow band that we consider. We ignore the Galactic absorption  because the Geminga is a quite nearby source ($d_{\rm pul}\sim 250$ pc). The upper limit flux $\Phi_{n_\sigma}$ is estimated with the Eq. (3) in \cite{Hollowood18}:
\begin{equation}
\Phi_{n_\sigma}=\Phi_{\rm model}\cdot\frac{n_{\sigma}\sqrt{N_{\rm obs}}}{N_{\rm model}}
\end{equation}
where $\Phi_{\rm model}$ is the model flux and $N_{\rm model}$ is the product of the exposure time and the model count rate, $n_\sigma=3$ in our case, and $N_{\rm obs}$ is the total observed number of cosmic background counts. The resultant $3\sigma$ upper limit flux are $f_{\rm cha,0.7-2.0\ {\rm keV}}=6.1\times 10^{-15}\ {\rm erg\ cm^{-2}\ s^{-1}}$ and $f_{\rm xmm,0.7-1.3\ {\rm keV}}=5.0\times 10^{-15}\ {\rm erg\ cm^{-2}\ s^{-1}}$ within $600''$ from \cha\ and \xmm\ data, respectively.

\begin{table*}[htbp]
 \centering
  \caption{X-ray observations}
  \begin{tabular}{@{}lcccc@{}}
\hline\hline
Name & \cha\ data & Exposure (ks) & \xmm\ data & Exposure (ks)\\ 
\hline
Geminga & 7592 14691 14692 14693 14694 & 659.3/660.7  &  0111170101 0201350101 0311591001  & 187.4/191.1/292.2$^a$  \\      
    & 15551 15552 15595 15622 15623 & & 0400260201 0400260301 0501270201 & \\
   & 16318 16319 16372 & & 0501270301 0550410201 0550410301 & \\
\hline
\end{tabular}
\begin{tablenotes}
\item
$^a$EPIC exposures of the clean MOS1/MOS2/total.
\end{tablenotes}
\label{t:obs}
\end{table*}

\section{Constraint on the Magnetic Field in the TeV halo}
The diffuse X-ray flux depends on the electron density and the strength of magnetic field. Since the former quantity can be evaluated through modelling HAWC's observation, we first need to find out the electron distribution in the TeV halo and fit the observed multi-TeV flux and SBP measured by HAWC. 

\subsection{Theoretical modelling}
{The Geminga pulsar has a high proper motion velocity of $211(d_{\rm pul}/250\,\rm pc) \,kms^{-1}$ \citep{Faherty07}, implying that the pulsar has moved $\sim 70$pc from its birthplace given an age of $\tau_{\rm pul}=342\,$kyr. However, the cooling timescale of $100\,$TeV electrons which are relevant for TeV and X-ray emission is several times $10^{11}\,$s in ISM. For the best-fit diffusion coefficient obtained in HAWC's paper ($4.5\times 10^{27}\rm \,cm^2s^{-1}$ at 100\,TeV), TeV- and X-ray emitting electrons diffuse a distance of $\sim 40\,$pc while the pulsar only moves $\sim 5\,$pc within the lifetime of these energetic electrons. We therefore neglect the proper motion of Geminga in this work for simplicity}, and deal with the electron transport in the spherical coordinate, defining the pulsar location (i.e., particle injection) at $r=0$ and assuming spherical symmetry for the particle transport.  

The differential density of electron with energy $E_e$ at a distance $r$ and at a time $t$ after the initial injection, i.e., $N(E_e, r, t)$, can be given by
\begin{equation}\label{eq:transport}
\frac{\partial N(E_e,r,t)}{\partial t}=\frac{1}{r^2}\frac{\partial}{\partial r}\left(r^2D(E_e,r)\frac{\partial N}{\partial r} \right)-\frac{\partial}{\partial E_e}\left(\dot{E}_eN\right)+Q(E_e, t)\delta(r),
\end{equation}
where $\delta(r)$ is the Dirac function. $D(E_e,r)$ is the diffusion coefficient at a distance $r$ from the pulsar, which is assumed to be
\begin{equation}\label{eq:D}
D=\left\{
\begin{array}{ll}
D_1, \quad r\leq r_0,\\
D_2, \quad r>r_0
\end{array}
\right.
\end{equation}
with $r_0$ being the radius of the boundary between the inner inefficient diffusion region and outer normal diffusion region.
$\dot{E_e}$ is the cooling rate of electrons due to synchrotron radiation and inverse Compton radiation, which can be given by \citep{Moderski05}
\begin{equation}
\dot{E}_e=-\frac{4}{3}\sigma_Tc\left(\frac{E_e}{m_ec^2}\right)^2\left[U_B+U_{\rm ph}/(1+4\frac{E_e\epsilon_0}{m_e^2c^4})^{3/2}\right]
\end{equation}
where $\sigma_T$ is the Thomson cross section, $m_e$ is the electron mass and $c$ is the speed of light. $U_B=B^2/8\pi$ is the magnetic field energy density with $B$ following the same form of the diffusion coefficient, i.e.,
\begin{equation}\label{eq:Bfield}
B=\left\{
\begin{array}{ll}
B_1, \quad r\leq r_0,\\
B_2, \quad r>r_0,
\end{array}
\right.
\end{equation}
and $U_{\rm ph}$ is the radiation field energy density. $\epsilon_0$ is the average photon energy of the radiation field which is equal to $2.82kT$ in the case of black body or grey body radiation with $k$ being the Boltzmann constant and $T$ being the temperature. Following \citet{HAWC17_Geminga}, in addition to the CMB, we also consider a 20\,K infrared photon field and a 5000\,K optical photon field as the background photon field. The spectrum of both two photon fields are assumed to follow a greybody distribution with an energy density of $0.3\,\rm eV/cm^3$, as approximately derived by GALPROP \citep{Moskalenko98}.
The injection spectrum of electrons at any given time $t$ is assumed in a form of power-law function with a high-energy cutoff, i.e., 
\begin{equation}
Q(E_e,t)=S(t)N_0E_e^{-p}e^{-E_e/E_{e, \rm max}}
\end{equation}
with $p$ being the spectral index and $E_{e,\rm max}$ being the maximum energy of electron injected by the pulsar, and $S(t)\propto 1/(1+t/\tau_s)^2$ assuming the pulsar is a pure dipole radiator of a braking index of 3, where $\tau_s$ is the spin-down timescale of the pulsar and $\tau_s=12\,$kyr is adopted. $N_0$ is the normalization and can be found by $\int\int dt dE_e Q(E_e,t)=W_e$.

In order not to introduce too many free parameters, we fix $D_2=D_{\rm ISM}=3.86\times 10^{28}(E_e/1\rm GeV)^{1/3}\rm cm^2s^{-1}$ and $B_2=3\,\mu$G in our calculation. The energy dependence of $D_1$ is also assumed to be $\propto E^{1/3}$ following the Kolmogorov-type turbulence, i.e., $D_1=D_1(100{\rm TeV})(E_e/100{\rm TeV})^{1/3}$, while the normalization, i.e. $D_1(100{\rm TeV})$, is allowed to change. On the other 
hand, we also require the value of $D_1$ at $E_{e,\rm max}$ to be larger than the Bohm diffusion coefficient ($D_B(E_{e,\rm max})=r_g(E_{e,\rm max})c/3$ where $r_g(E_{e,\rm max})=E_{e,\rm max}/eB$ is the Larmor radius of the maximum energy electron) which is the limiting case for the diffusion coefficient, so $D_1$ is not a totally free parameter.
Since \cite{Xi18} shows that a hard injection spectrum is required to be consistent with the upper limit of multi-GeV flux from the observation of {\it Fermi}-LAT, we fix $p=1.6$ for the moment. Actually, the spectral index is not important in this work. As we mentioned before, the energies of X-ray emitting electrons and TeV emitting electrons are quite close, so the expected X-ray flux is insensitive to the spectral index $p$ as long as HAWC's observation is fitted (see Section 3.3). Note that the photon index of the TeV emission measured by HAWC is -2.34. Given a hard electron spectrum $p=1.6$, we need a relatively small cutoff in the injection electron spectrum to reproduce the observed spectrum, and therefore $E_{e,\rm max}=100-200$\,TeV is employed. {Such a high energy in principle can be achieved in the pulsar wind termination shock although some details on the acceleration mechanism remain unclear \citep[see][and references therein]{Kirk09, Aharonian13}.} The minimum electron energy at injection is fixed at 1\,GeV. Then, the free parameters left in our calculation are the total injection energy of electrons $W_e$, the magnetic field in the TeV halo $B_1$ and the halo diffusion coefficient $D_1(100\rm TeV)$.

\begin{figure}
\includegraphics[width=1\columnwidth]{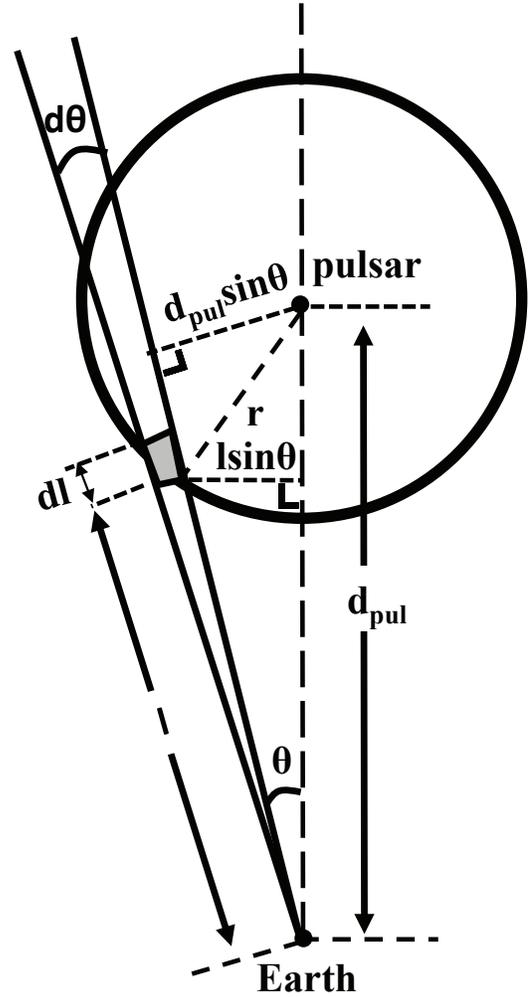}
\caption{Diagrammatic sketch for calculating the integrated flux from the TeV halo. The system is symmetric with respect to the axis connecting Earth and the pulsar. The shaded region is the element volume in the integration. See Section 3.1 for details.}\label{fig:sketch}
\end{figure}

We then solve the equation numerically, from $t=0$ when the pulsar just starts to inject electrons to the current age of the pulsar $\tau_{\rm pul}$, by discretizing the equation into second-order accuracy in both space and energy dimension based on the method introduced in Appendix.~\ref{sec:append}. After obtaining the electron distribution, we can calculate the SBP from the radiation of the electrons. We focus on the contribution of electrons within a sphere of a sufficient large radius $r_{\rm max}$\footnote{We have tried the calculation with $r_{\rm max}=150$, 200, 249, 249.99\,pc, the results are almost the same.}. For certain viewing angle $\theta$ from the pulsar's position, we integrate the emission of electrons in the line of sight, say, from a minimum distance of $l_{\rm min}=d_{\rm pul}\cos\theta-\sqrt{r_{\rm max}^2-d_{\rm pul}^2\sin^2\theta}$ to Earth to a maximum distance of $l_{\rm max}=d_{\rm pul}\cos\theta+\sqrt{r_{\rm max}^2-d_{\rm pul}^2\sin^2\theta}$ to Earth (see Fig.~\ref{fig:sketch}). At a given point in the line of sight with a distance $l$ to Earth ($l_{\rm min}\leq l\leq l_{\rm max}$, the radius of this point from the pulsar $r$ can be found by $r=\sqrt{l^2+d_{\rm pul}^2-2ld_{\rm pul}\cos\theta}$ where $d_{\rm pul}$ is the distance between the pulsar and Earth, and the electron density can be obtained via interpolation based on the obtained $N(E_e,r,\tau_{\rm pul})$ where $\tau_{\rm pul}$ is the age of the pulsar at the present time. Thus, the element volume in the neighbourhood of this point can be given by $dV=2\pi l\sin\theta \cdot ld\theta \cdot dl$ where $2\pi$ considers the symmetry of the system. Let us define an operator $\mathcal{F}$ which  calculates the differential spectrum of synchrotron radiation and IC radiation of electrons $F(\epsilon)$, following the formulae in \citet{Rybicki79}, given the electron density, the magnetic field or the background photon field. The flux emitted by electrons in the element volume can then be obtained by
\begin{equation}
dF(\epsilon)=\mathcal{F}\left\{N[E_e,r(l),\tau_{\rm pul}], B[r(l)], T_{\rm ph}, U_{\rm ph}  \right\}dV/4\pi l^2.
\end{equation}
Thus, the total photon flux from an annular region between $\theta$ and $\theta+d\theta$ centred at the pulsar in the celestial sphere can be given by
\begin{equation}
I(\epsilon, \theta)\cdot 2\pi \sin\theta d\theta=\int_{l_{\rm min}}^{l_{\rm max}}\mathcal{F}\left\{N[E_e,r(l),\tau_{\rm pul}], B[r(l)], T_{\rm ph}, U_{\rm ph}  \right\}dV/4\pi l^2
\end{equation}
where $I(\epsilon,\theta)$ is the intensity of the annular region. Finally, we arrive at
\begin{equation}
I(\epsilon,\theta)=\frac{1}{4\pi}\int_{l_{\rm min}}^{l_{\rm max}}dl\mathcal{F}\left\{N[E_e,r(l),\tau_{\rm pul}], B[r(l)], T_{\rm ph}, U_{\rm ph}  \right\}
\end{equation} 
The total flux within certain angle $\theta_0$ from the pulsar can be obtained by $F(\epsilon)=2\pi\int_0^{\theta_0}I(\epsilon,\theta)\sin\theta d\theta$.

\subsection{Results}
\begin{figure}[htbp]
\centering
\includegraphics[width=1\columnwidth]{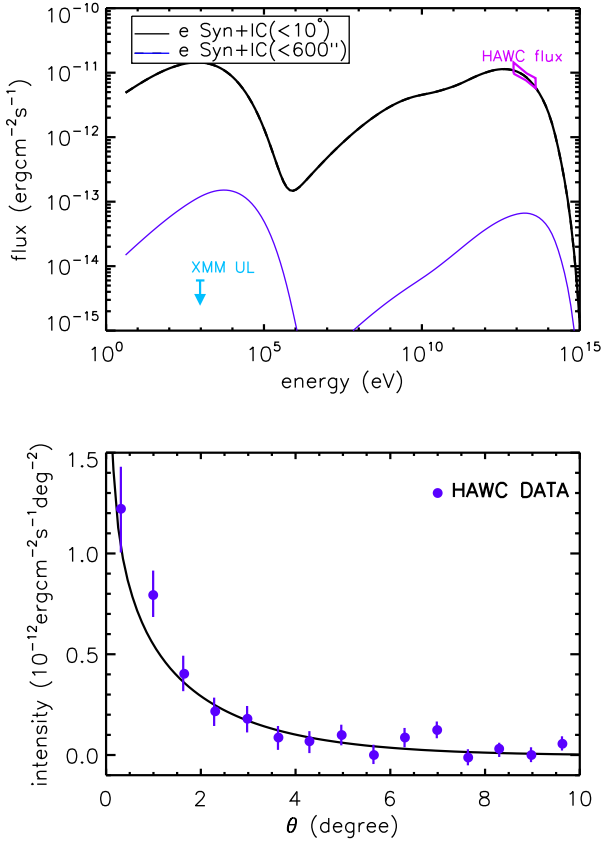}
\caption{{\bf Upper panel}: Expected multiwavelength flux from the TeV halo of Geminga produced by the synchrotron and IC radiation of injected electrons, in comparison with data from various wavelength. The black solid curve represent the expected flux from a region of $10^\circ$ around Geminga. The magenta bowtie shows the measured flux by HAWC from the same region \citep{HAWC17_Geminga}. The blue solid curve represents the integrated flux from a region of 600'' around Geminga and the cyan arrow is the upper limit of \xmm\  for the same region obtained in this work. The upper limit of \cha\ almost overlap with the \xmm\ upper limit so we do not show it in the figure for clarity. {\bf Lower panel:} the corresponding SBP of 8-40\,TeV emission. Employed parameters: $B_1=3\,\mu$G, $D_1(100{\rm TeV})=4.5\times 10^{27}\rm \, cm^2/s$, $W_e=4.2\times 10^{47}\,$ergs, $r_0=50\,$pc, $E_{e,\rm max}=200\,$TeV and $p=1.6$.}\label{fig:highB}
\end{figure}

We now apply the method to the TeV haloes of Geminga ($d_{\rm Gem}=250$pc, $\tau_{\rm Gem}=342\,$kyr). Firstly, as an example, we show in the upper panel of Fig.~\ref{fig:highB} the multiwavelength flux from a region within $10^\circ$ of Geminga (the black solid curve) and within $600''$ of Geminga (the blue solid curve) in the celestial sphere, respectively. The magnetic field and the diffusion coefficient for the TeV halo are the same with best-fit parameters obtained in \citep[][i.e., $B_1=3\mu$G and $D_1(100{\rm TeV})=4.5\times 10^{27}\rm \, cm^2/s$]{HAWC17_Geminga}, and other parameters can be found in the caption of Fig.~\ref{fig:highB}. The theoretical SBP (the black solid curve) is compared to the observed one in the lower panel. As is shown, while the HAWC's observation is explained, the associated X-ray flux overshoots the upper limits of \xmm\ and \cha\footnote{We only show the \xmm\ upper limit in the figure for clarity, as the upper limits of these two instruments are basically the same.} by one order of magnitude. 

\begin{figure*}[htbp]
\includegraphics[width=1\columnwidth]{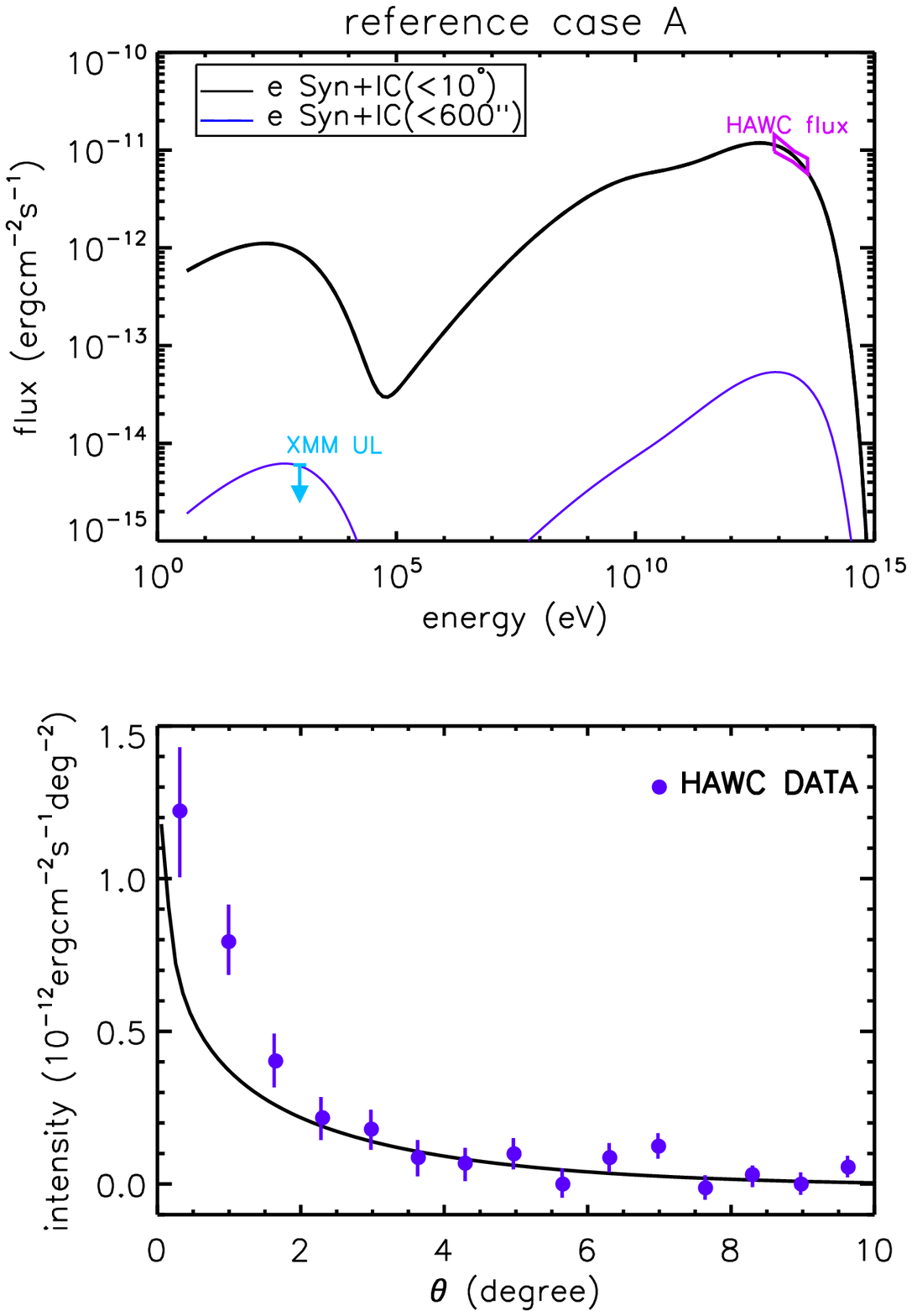}
\includegraphics[width=1\columnwidth]{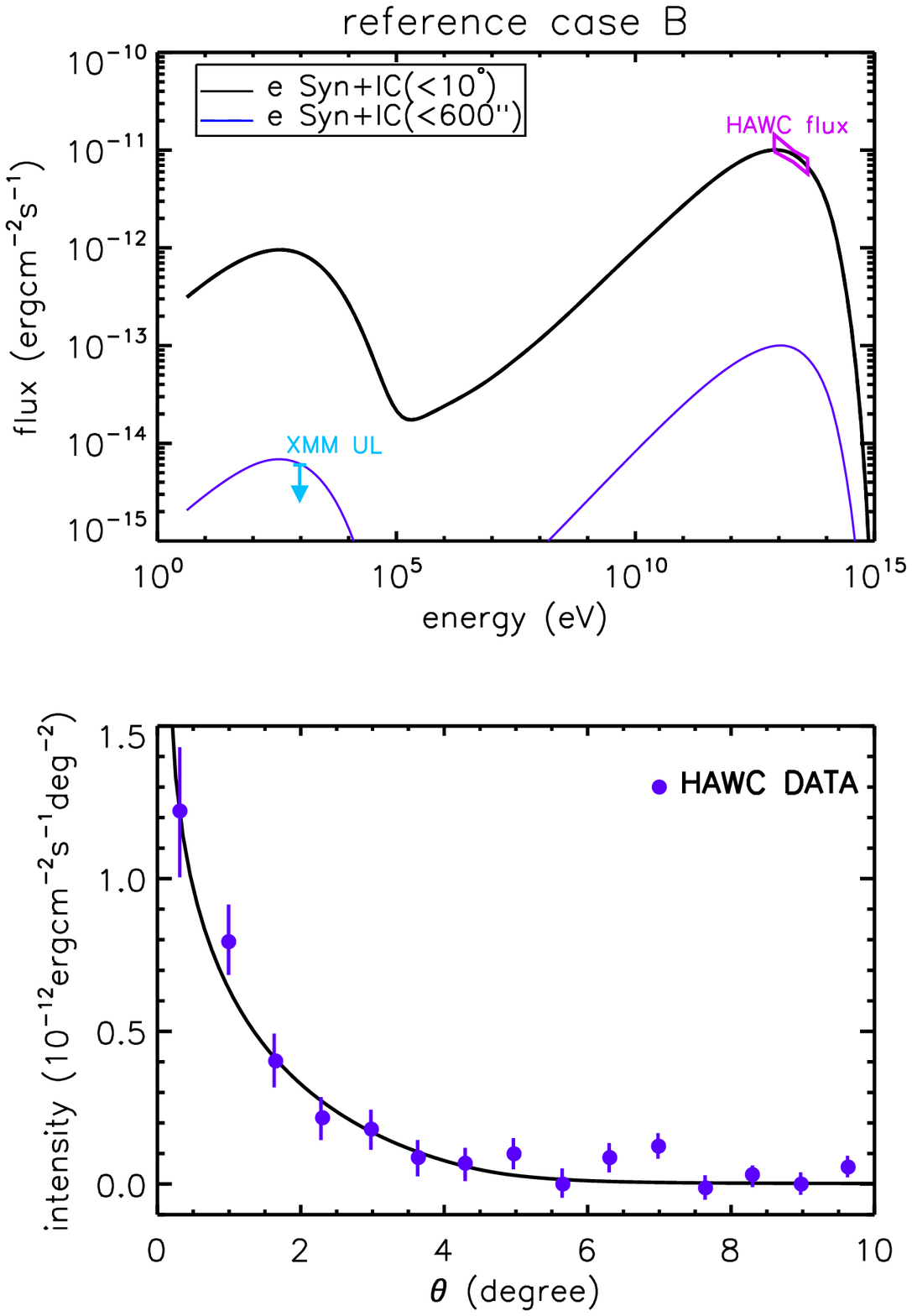}
\caption{Same as Fig.~\ref{fig:highB}, but for smaller magnetic field in the vicinity of Geminga ($B_1$). {\bf Left: (reference case A)} results under two-diffusion zones with $B_1=0.8\,\mu$G, $D_1(100\,{\rm TeV})=D_B({100\,\rm TeV})=4.2\times 10^{27}\rm cm^2s^{-1}$, $W_e=2.9\times 10^{47}$ergs; {\bf Right: (reference case B)} results under a continuously changing diffusion coefficient with $B_1=0.6\,\mu$G, $D_1(100\,\rm TeV)=D_B(100\,\rm TeV)=5.6\times 10^{27}cm^2s^{-1}$, $W_e=7.7\times 10^{47}$erg. In both left and right panels, the maximum electron energy is fixed at 100\,TeV}\label{fig:lowB}
\end{figure*}

We find that $B_1$ needs to be tuned down to $\sim 0.8\,\mu$G or smaller to be consistent with the upper limit of X-ray flux. For such a small magnetic field the cooling becomes inefficient. The cooling timescale of electrons $t_c=-E_e/\dot{E}_e$ becomes longer than that in the case of $B_1=3\mu$G, allowing $100\,$TeV electrons transport to a farther distance before being cooled. Consequently, the SBP of TeV emission would become too flat compared to the observed one. To make the SBP profile as steep as the observation, a smaller diffusion coefficient is required to keep more electrons closer to the pulsar, by making $\sqrt{D_1t_c}$ roughly remain the same as that in the reference case. For $100\,$TeV electrons, the cooling time with $B_1=3\,\mu$G is $t_c=3\times 10^{11}$s, while $t_c=9\times 10^{11}$s for $B_1=0.8\,\mu$G (note that in this case the IC process dominates the cooling). Thus, it requires $D_1(100{\rm TeV})\simeq 1.6\times 10^{27}\,\rm cm^2/s$, which is, however, smaller than the Bohm diffusion coefficient for $B_1=0.8\,\mu$G above a few tens of TeV, i.e.,$D_B=4.2\times 10^{27}(E_e/100{\rm TeV})\,\rm cm^2/s$. We try to fit the SBP with physically reasonable diffusion coefficient by normalizing $D_1(100\rm TeV)$ to $D_B(100\rm TeV)$. The results are shown in the bottom left panel of Fig.~\ref{fig:lowB} and we can see the SBP is too flat compared to the observation (i.e., the intensity at small angle is lower than the measured one). In the following, we denote this case by reference case A.


\citet{Lopez18} argued that a too weak magnetic field ($B<2\mu$G) can be ruled out according to the bad fitting to the SBP. However, if we relax the condition for a spatially constant diffusion coefficient inside the TeV halo, the fitting can be improved by considering a continuously decreasing diffusion coefficient with $r$ in the TeV halo within $r_0$. Such a kind of diffusion coefficient might be possible given that the CR self-confinement will be less efficient as the ion-neutral damping of the generated waves can be more important at larger distance to the pulsar \citep[e.g.][]{Evoli18}. In this case, electrons diffuse faster as they propagate to larger radius and hence the electron distribution will show a larger negative gradient outwards. Phenomenologically, there are multiple ways of constituting a function for $D(r)$ to give a consistent result with the observed morphology. As an example, we find the following form for $D$, i.e.,
\begin{equation}\label{eq:continuousD}
D=\left\{
\begin{array}{lll}
D_1, \quad r<20{\rm pc}\\
D_1\left(\frac{D_2}{D_1}\right)^{(r-20)/30}, \quad 20{\rm pc}\leq r < 50{\rm pc}\\
D_2=D_{\rm ISM}, \quad r\geq 50{\rm pc}
\end{array}
\right.
\end{equation}
with $D_1=5.6\times 10^{27}(E_e/100{\rm TeV})^{1/3}\rm cm^2s^{-1}$ and $B_1=0.6\,\mu$G can give a reasonable fitting to the data. Here $5.6\times 10^{27}\rm cm^2s^{-1}$ is the Bohm diffusion coefficient at 100\,TeV under a magnetic field of $0.6\mu$G. The logarithm of the diffusion coefficient decreases linearly with $r$ from $D_1$ at $r=20$\,pc to $D_2$ at $r=50\,$pc. The results are shown in right panels of Fig~\ref{fig:lowB}. Note that an even smaller magnetic field is required inside the TeV halo to be consistent with the X-ray upper limit because of a steeper gradient of electrons density distribution is obtained in this case. As a consequence, when the flux of TeV emission is fitted, there are more electrons distributing at small radius than that in the reference case A, leading to a higher X-ray flux at the small radius. A weaker magnetic field $B_1$ is thus needed to reduce the expected X-ray flux. We denote this case by reference case B.

\subsection{Influence of some model parameters}
In this section, we discuss the influences of some model parameters and demonstrate that the requirement of a weak magnetic field in the TeV halo of Geminga is robust.

\begin{figure*}
\includegraphics[width=0.35\textwidth]{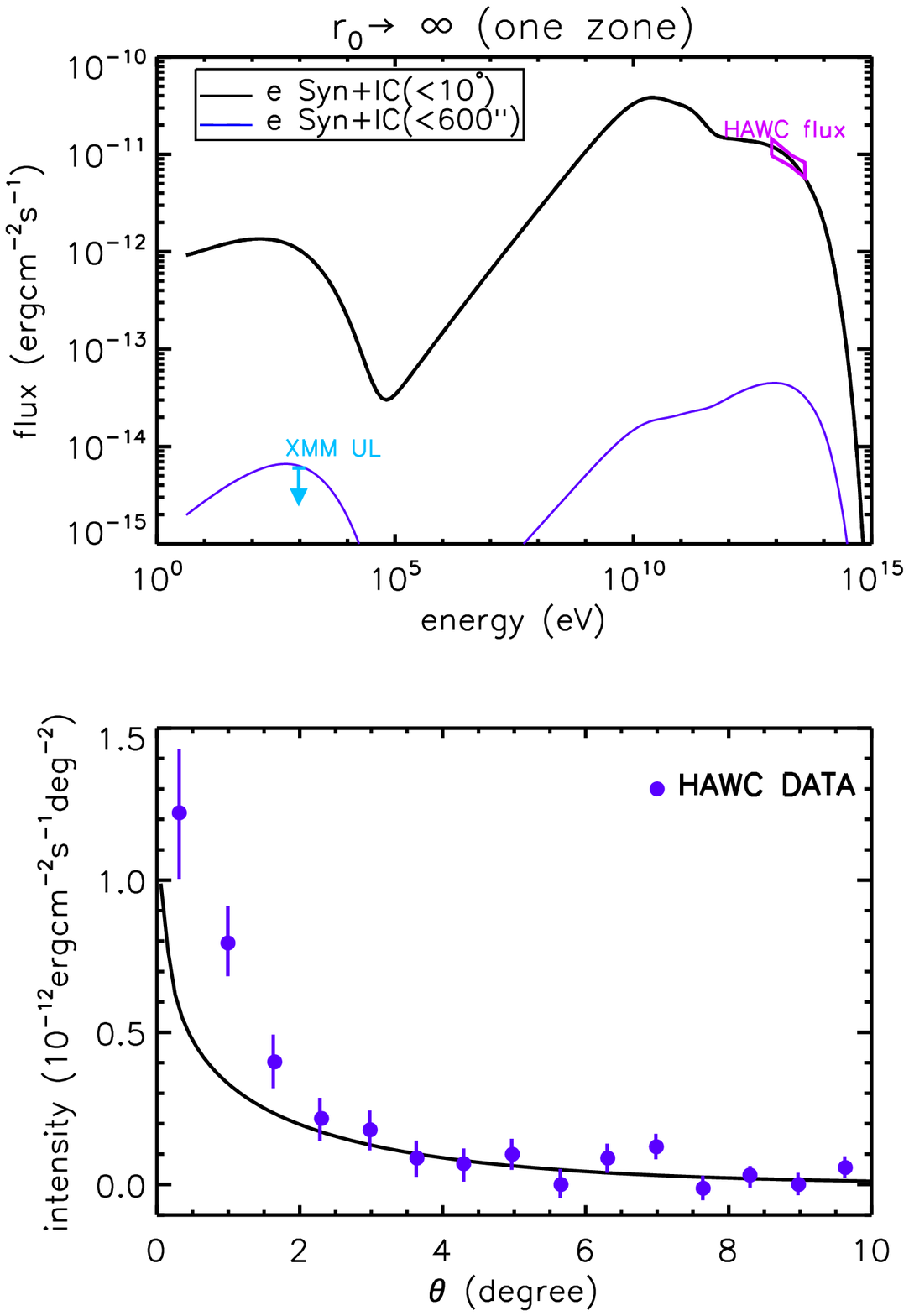}
\includegraphics[width=0.35\textwidth]{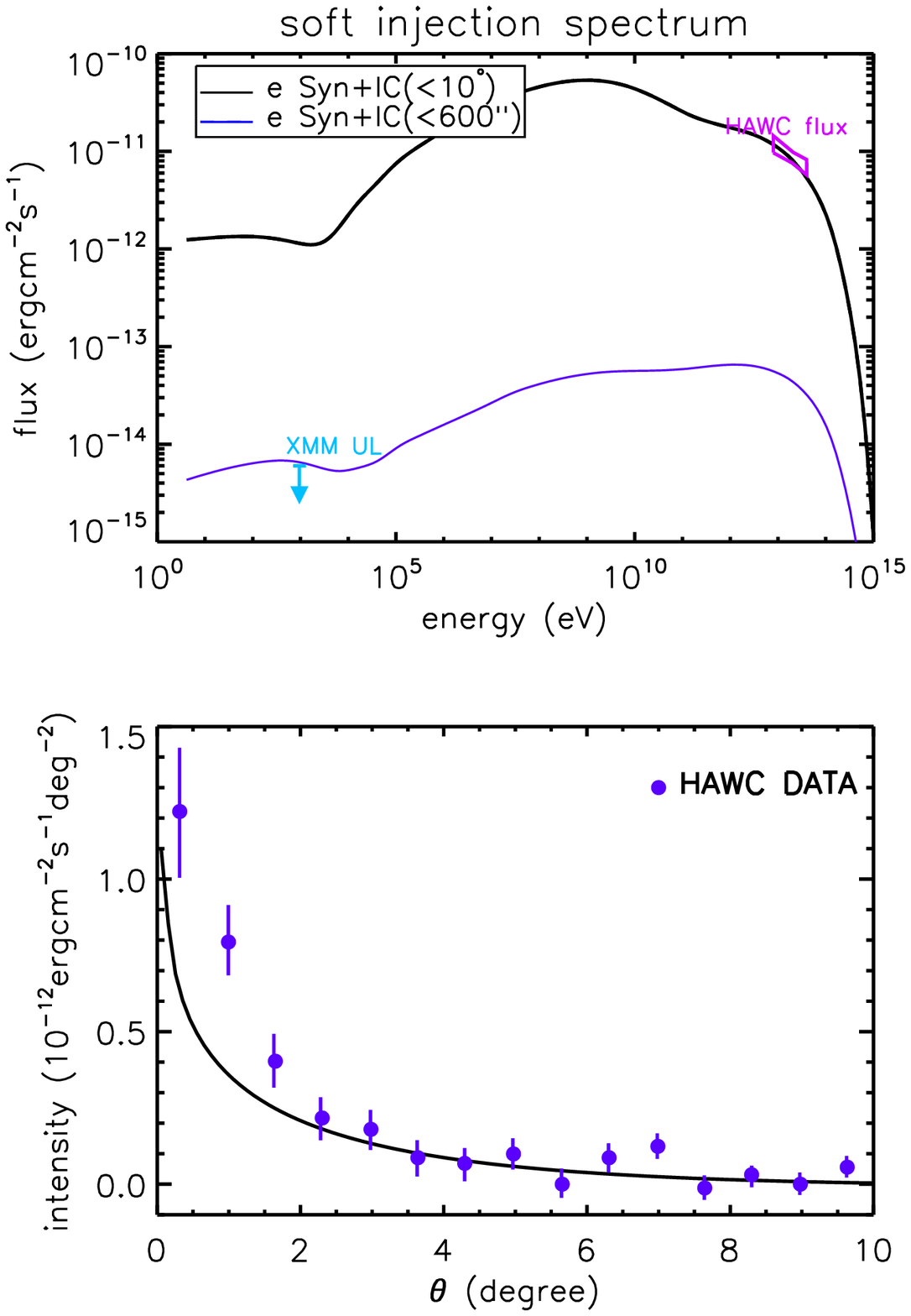}
\includegraphics[width=0.35\textwidth]{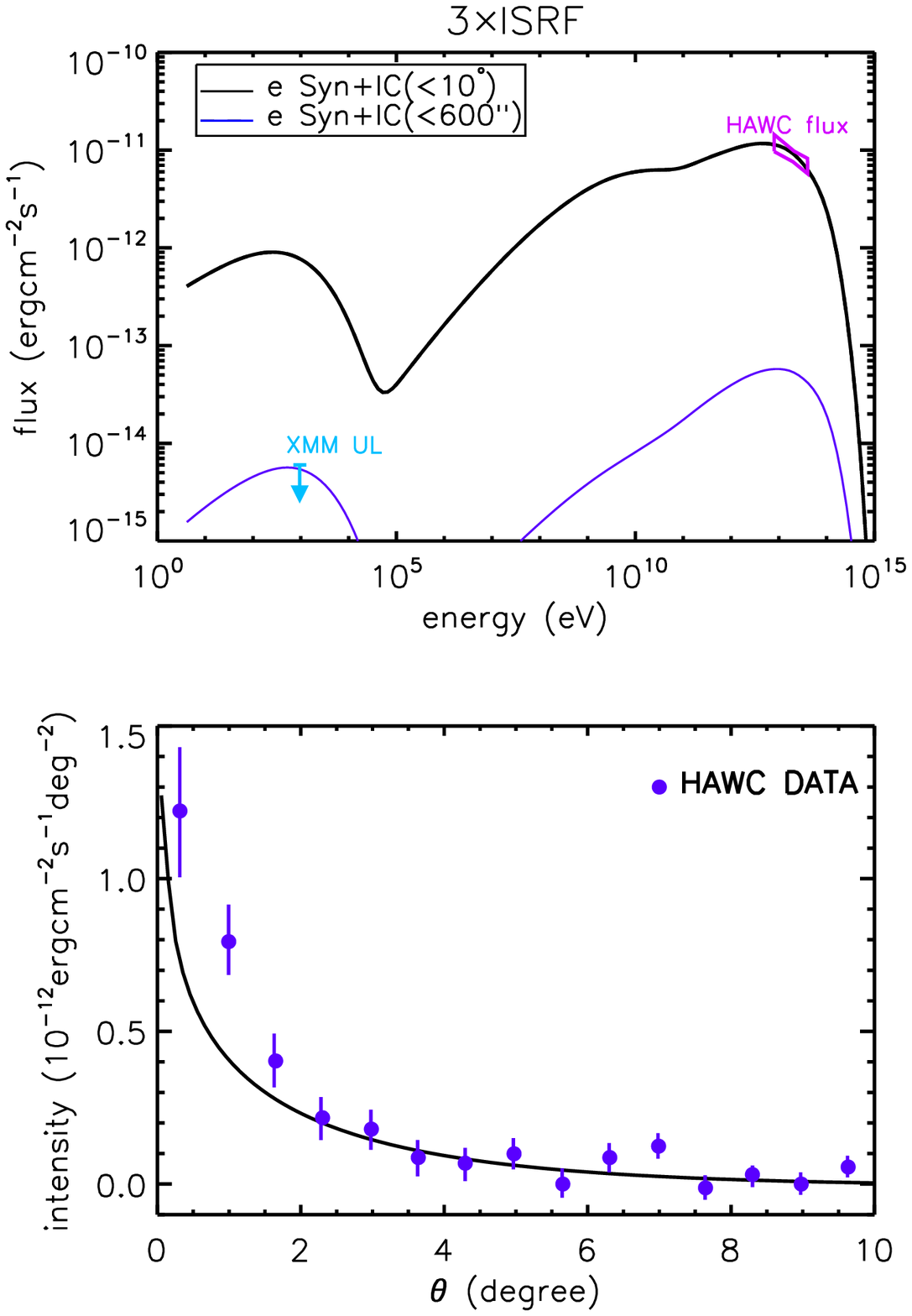}
\caption{Influence of model parameters. {\bf Left:} the case of $r_0\to \infty$ (one zone). $B_1=0.9\mu$G, $W_e=2.7\times 10^{47}\,$ergs; {\bf Middle:} the case of a soft injection spectrum for electrons with $p=2.2$. $B_1=0.9\mu$G, $E_{e,\rm max}=200\,$TeV, $W_e=3.3\times 10^{48}$ergs; {\bf Right:} the case of three times higher density for the interstellar radiation field. $B_1=0.9\mu$G and $W_e=2.4\times 10^{47}\,$ergs. For all three cases, unlisted parameters are the same as those in the reference case A. See Section~3.3 for more discussions.}\label{fig:compare}
\end{figure*}

First, we discuss the influence of $r_0$, i.e., the radius of the boundary between the inner inefficient diffusion zone and the outer standard diffusion zone. HAWC's measurement on the SBP suggests an inefficient diffusion zone extending to at least $\sim 30$\,pc away from the Geminga pulsar. Thus, one can in principle assume $r_0$ to be any value $>30\,$pc. In the reference cases, we fix $r_0$ to be 50\,pc and a typical ISM magnetic field $B_2=B_{\rm ISM}=3\mu$G is assumed for $r>r_0$. A smaller $r_0$ cannot help to avoid a weak magnetic field in the vicinity of the pulsar, because a steeper gradient of the electron distribution will be obtained for a smaller $r_0$. It will only result in a stronger X-ray flux from the small radius where the X-ray upper limit is extracted from so that an even weaker magnetic field has to be assumed for $r<r_0$. This is also the reason why a weaker magnetic field is required in the reference case B than in the reference case A as discussed above. On the other hand, for a larger $r_0$ the weak magnetic field region will occupy a larger volume and hence more electrons will radiate in the weak magnetic {field} $B_1$. We thus question ourself whether the value of $B_1$ can be increased to certain extent in this case. We then consider a limiting case of $r_0\to \infty$, which can be also regarded as only one diffusion zone.  Despite the {\it Fermi}-LAT upper limit and the flat SBP due to inefficient cooling and inefficient diffusion of electrons, the value of $B_1$ can be increased, however, only to $0.9\mu$G (see left panels of Fig.~\ref{fig:compare}). This is because the contribution of electrons far from the pulsar is subdominant as the electron density is quite low at large $r$, and therefore assuming a weak magnetic field for a much larger region does not reduce the synchrotron flux significantly. 

In both two reference cases, we fix the spectral index for injection electrons at $p=1.6$ and argue that the result is not sensitive to the injection spectral index, since the X-ray-emitting electrons and the TeV-emitting electrons are the same and thus the amount of the X-ray-emitting electrons can be normalized by the HAWC's observation. While this argument is generally true, more precisely, the average energy of X-ray-emitting electrons is a little higher than that of the TeV-emitting electrons especially given a weak magnetic field. One may then wonder that whether a soft injection spectrum could reduce the X-ray flux and increase $B_1$ to certain extent. We here consider an injection spectral index of $p=2.2$. Again, regardless of the {\it Fermi}-LAT upper limit and the flat SBP, we find that while TeV flux is reproduced, a magnetic field of $B_1\leq 0.9\mu$G,  a little higher compared to that in reference cases, is needed in this case (see middle panels of Fig.~\ref{fig:compare}), which is, however, still significantly smaller than the typical ISM magnetic field. Although an even softer injection spectrum ($p>2.2$) can further allow a larger $B_1$, the resulting TeV spectrum will be too soft to be consistent with the HAWC's observation. From this figure, we can also know that the energy dependence $\delta$ of the diffusion coefficient cannot significantly influence the result neither, as the energy of X-ray-emitting electrons and the energy of TeV-emitting electrons are very close.

IC radiation competes with synchrotron radiation. The peak flux of the former radiation process to the latter one is roughly proportional to the ratio of the background photon density to the magnetic field density. Although there are uncertainties in the interstellar infrared photon field and optical photon field, these two background photon fields actually are not very relevant to the X-ray-emitting and TeV-emitting electrons due to the Klein-Nishina effect. Nevertheless, we artificially increase the energy densities of the infrared photon field and the optical photon field by a factor of 3 to see the possible influence. As is shown in the right panels of Fig.~\ref{fig:compare}, the magnetic field for the inefficient diffusion zone (i.e., $B_1$) can be increased slightly to about $0.9\mu$G in this case.

\section{Discussion}

\subsection{Low magnetic field in the vicinity of the pulsar}
\citet{Mattana09} found that the ratio between the gamma-ray (1-30\,TeV) and the X-ray (2-10\,keV) flux of TeV PWNe and PWNe candidates detected by H.E.S.S. increases with the characteristic age of the parent pulsars. This discovery is consistent with our result here, given an age of 342\,kyrs for Geminga. The authors explained this empirical relation as the different cooling time for X-ray emitting electrons (cooled) versus TeV-gamma-ray-emitting electrons (uncooled), which is, however, not applicable to the case of Geminga due to two reasons. First, the obtained X-ray upper limit is in the range of 0.7-1.3\,keV and the gamma-ray emission measured by HAWC is in the range of $8-40\,$TeV, which arise from electrons with quite close energies via synchrotron radiation and IC radiation respectively. What's more, even if the energy of gamma-ray radiating electrons is smaller than that of the X-ray emitting electrons, the age of Geminga pulsar is long enough to make gamma-ray-emitting electrons cooled as well. Thus, the null detection of diffuse X-ray emission from the TeV halo results from a low magnetic field. Note that, although the X-ray observation only focuses on a very small region of a projected size of $\lesssim 1\,$pc, the measured flux is contributed by all the electrons in the line of sight and hence the size of the low magnetic field region is required to be at least a few tens of parsecs. The obtained upper limit for the magnetic field in the vicinity of Geminga is much weaker than the typical ISM magnetic field. This may imply that the PWN of Geminga has experienced a significant expansion and its size is much larger than the nebula observed in X-ray which is $<0.1\,$pc \citep{Caraveo03}. A relatively weak magnetic field is also inferred in another TeV PWN, HESS~J1825-137 \citep{HESS06_PWN}, with a size of a few tens of parsecs. Since the TeV luminosity of this PWN is higher than the current spin-down luminosity of the parent pulsar, a natural explanation of the origin of the TeV nebula would then be the ``relic'' multi-TeV electrons injected in the past when the spin-down power of the pulsar was much higher. The survival of the multi-TeV electrons requires the magnetic field not to significantly exceed $1\mu$G\citep{HESS06_PWN, Aharonian13}. 

The low diffusion coefficient resulting from HAWC's observation together with the low magnetic field suggests a highly turbulent magnetic field in the TeV halo with field perturbation $\delta B/B$ of unity. Plasma instabilities, such as the streaming instability driven by CR gradient, must grow efficiently to saturation for wave number $k<2\pi/r_g(100\rm TeV)$. \citet{Quenby18} and \citet{Evoli18} suggested that CR self-regulation can be important around Geminga. A self-consistent study including the growth of the streaming instability, its feedback on the CR transport as well as modelling the multiwavelength emission is needed to justify this possibility.

\subsection{A Scenario of Hadronic Origin of the TeV emission}
Since pulsars and PWNe have been suggested as accelerators of cosmic ray protons \citep[e.g.][]{Cheng86, Gallant99, Arons03, Lemoine15}, we now investigate whether such a weak magnetic field can be avoided if we ascribe the TeV halo to the $pp$ collision between the accelerated protons and the surrounding ISM, while the electron injection from the pulsar can be very low in this case. Owing to the inefficient cooling of protons, historically injected protons can survive to the present day. Since the spin-down power of the pulsar was higher in the past, the accumulated protons might provide sufficient energy budget.

The gas density in the very vicinity of Geminga is about $n_g\sim 0.1\,\rm cm^{-3}$, as inferred from the equilibrium between the ram pressure of the pulsar wind and the thermal pressure of the ISM \citep{Caraveo03}. Such a low gas density results in a low energy loss rate of proton via the $pp$ collision, i.e., $\frac{1}{E_p}\frac{dE_p}{dt}\simeq 0.17\sigma_{pp}n_gc=2\times 10^{-17}\rm s^{-1}$. Given that the $8-40$\,TeV luminosity within $10^\circ$ of Geminga is about $L_{\rm TeV}=10^{32}\rm \,ergs^{-1}$, it requires the energy of $\sim 80-400$\,TeV CR protons inside the TeV halo to be $L_{\rm TeV}/(\frac{1}{E_p}\frac{dE_p}{dt})=5\times 10^{48}$erg. Such an amount of CR proton energy is only a fraction of the total spin-down power of Geminga (which is a few times $10^{49}$ergs). On the other hand, a continuously decreasing diffusion coefficient with $r$ similar to that in the reference case B is also needed in the hadronic interpretation, as otherwise the expected SBP would be too flat to be consistent with the observed one given the inefficient cooling. However, lots of protons will then escape to much farther distances and a huge amount of injection proton energy is consequently required.

\begin{figure}[htbp]
\includegraphics[width=1\columnwidth]{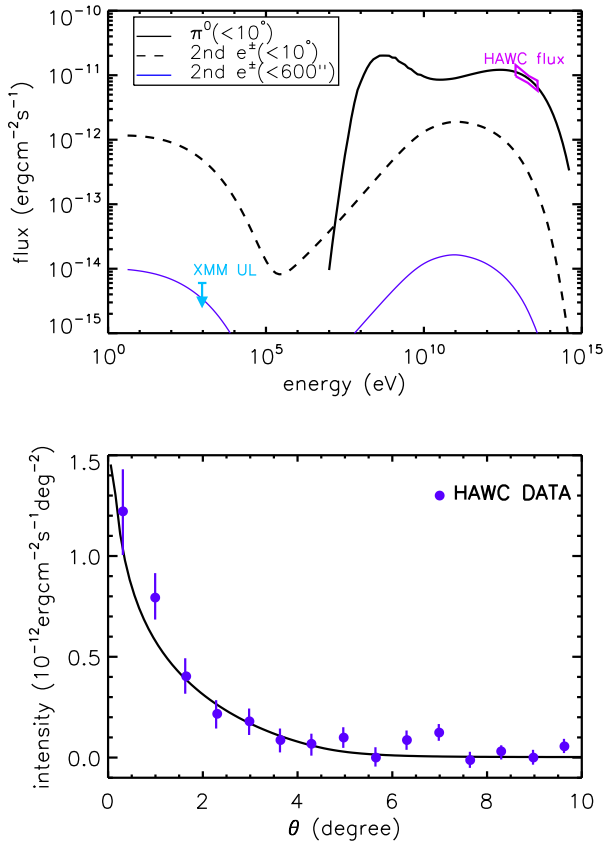}
\caption{The expected multiwavelength flux ({\bf upper panel}) and SBP in the hadronic model ({\bf lower panel}). The black solid curve is the pionic emission arising from $pp$ collision. The synchrotron and IC emission from secondary electrons within $10^\circ$ of Geminga (black dashed curve) and within $600''$ of Geminga (blue curve) are also shown. The gas density is $n_g=0.1\rm cm^{-3}$ and the required total proton energy to fit the TeV flux is unreasonably high, i.e., $W_p=1.6\times10^{52}$ergs. See Section~4.2 for more discussion.}\label{fig:had}
\end{figure}

In Fig.~\ref{fig:had}, we show the expected multiwavelength flux and the SBP in the hadronic scenario, with $B_1=B_2=3\,\mu$G. The maximum injection proton energy is assumed to be $500\,$TeV. The diffusion coefficient follows the form of Eq.~\ref{eq:continuousD} with $D_1=3.3\times10^{27}(E_p/100{\rm TeV})^{1/3}\rm cm^2s^{-1}$, through the normalization $D_1(500{\rm TeV})=D_B(500\rm TeV)$ for $B_1=3\mu$G. With this diffusion coefficient, we find that the proton energy at 100\,TeV inside the inner diffusion zone is only $\sim 0.1$\% of total injected energy at 100\,TeV. As a result, the required total proton injection energy is found to be $\sim 10^{52}(n_g/0.1\rm cm^{-3})^{-1}$ergs, which is much larger than the spin-down energy of Geminga and the CR energy that a supernova remnant can supply, even for the typical ISM density of $n_g=1\rm cm^{-3}$. The proton injection energy can be reduced if more protons are injected at late time, such as with a constant injection luminosity (i.e., $S(t)\propto t^0$). However, the required proton injection energy in this case is still $\sim 10^{50}(n_g/1\rm cm^{-3})^{-1}$ergs. {Moreover, taking in account the proper motion of Geminga will lead to a larger demand on the energy budget, since the proton distribution will be more extended.}
One may employ an even smaller diffusion coefficient (on the order of $10^{26}\,\rm cm^2s^{-1}$ at 100\,TeV) to further reduce the required proton energy. However, a larger magnetic field has to be invoked in this case to ensure that the employed diffusion coefficient is larger than that in the Bohm limit. Note that we do not consider electron injection in this case, but there are still secondary electrons produced in the $pp$ collisions accompanying with gamma rays. As is shown with the blue curve in the upper panel of Fig.~\ref{fig:had}, the synchrotron radiation of secondary electrons is only marginally consistent with the X-ray upper limit with $B_1=3\mu$G. A larger magnetic field will result in a higher synchrotron flux of the secondary electrons and contradict with the X-ray upper limit. Thus, we conclude that a simple hadronic model is not a preferable solution.

\subsection{A scenario of an ordered magnetic field inside the TeV halo}
Before we make our conclusion, we would like to qualitatively discuss another possible scenario for the null detection of diffuse X-ray emission, in which the need for a weak magnetic field may be avoided. Let us denote the pitch angle between an electron and the magnetic field by $\alpha$. Due to the relativistic beaming effect, the synchrotron radiation of the electron with $E_e$ will be concentrated to a very small angle $1/(E_e/m_ec^2)$ of a cone with a half-opening angle $\alpha$. On the other hand, the synchrotron radiation power is $P_{\rm syn}\propto (B\sin\alpha)^2$. Thus, if the magnetic field in the TeV halo of Geminga is roughly aligned with (or oppositely aligned with) our line of sight to the pulsar, the synchrotron radiation of electrons that move towards us will be very weak given $\sin^2\alpha \ll 1$, while we cannot see the efficient synchrotron radiation of electrons with a large pitch angle since their radiation will be beamed into other direction. 


Note that the obtained results in previous sections are based on isotropic particle diffusion. However, electrons would experience anisotropic diffusion in the presence of such a mean magnetic field orientation in the TeV halo. Particles will diffuse faster in the direction parallel to the mean field than in the direction perpendicular to the mean field. The TeV SBP may still be explained since it reflects the projected electron distribution which are mainly relevant with the perpendicular diffusion. Interestingly, the expected positron flux at Earth contributed by Geminga  can be much higher than that expected in the isotropic diffusion scenario. A detailed modelling in this scenario will be helpful for verification. Such a calculation is not available with our current code, but will be an interesting project in the future.

\section{Conclusion}
In this work, we analysed the data of \xmm\ and \cha\ in a region of 600'' around the Geminga pulsar. No significant X-ray emission is detected, yielding an upper limit of $(5-6)\times 10^{-15}\,\rm ergcm^{-2}s^{-1}$ in $0.7-2.0$keV for the diffuse X-ray flux. We then modelled the TeV emission measured by HAWC under the constraint of the X-ray upper limit. By solving the 1D transport equation of injected electrons, we obtained the spatial distribution of electrons and then calculated the expected multiwavelength flux of electron from the TeV halo via the synchrotron radiation and inverse Compton radiation. On the premise of fitting the TeV emission, we found that the magnetic field in the TeV halo needs to be smaller than $0.8\mu$G in order not to overshoot the X-ray upper limit. The low magnetic field may imply that the pulsar wind nebula of Geminga has experienced significant expansion to a size much larger than the $<0.1$pc nebula observed in X-ray.  The weak magnetic field together with the small diffusion coefficient inferred from HAWC's observation implies that the Bohm diffusion may probably have achieved in the TeV halo. We also sought for alternative explanations for the null detection of diffuse X-ray emission without invoking a weak magnetic field and/or a small diffusion coefficient. We found that the hadronic interpretation of Geminga's TeV halo does not work, since it requires extreme parameters such as a huge amount of proton injection energy and/or very small diffusion coefficient. On the other hand, both the weak magnetic field and the small (parallel) diffusion coefficient may be avoided if the magnetic field in the TeV halo has a mean direction roughly (oppositely) aligned with our line of sight. 

\acknowledgements 
We thank Huirong Yan and Zhiyuan Li for helpful discussion. CG acknowledges support from the National Natural Science Foundation of China No. 11703090. X.-Y. W. is supported by the National Key R \& D program of China under the grant No.2018YFA0404200 and the NSFC grants No.11625312 and No.11851304.

\appendix
\section{Discretizing the particle transport equation}\label{sec:append}
To solve Eq.~(\ref{eq:transport}), we employ the operator splitting technique to simplify the problem into solving a parabolic partial differential equation (the diffusion in space) and a hyperbolic partial differential equation (the convection in energy space). The classical Strang splitting scheme which is of second order accuracy is adopted so that when going from $l$th to $l+1$th time step with the step length $\Delta t$, we have
\begin{equation}
N^{l+1}(r,E_e)=N^l(r,E_e)e^{\mathcal{L}_E\Delta t/2}e^{\mathcal{L}_r\Delta t}e^{\mathcal{L}_E\Delta t/2}
\end{equation}
where $\mathcal{L}_E$ and $\mathcal{L}_r$ are the differential operator for energy $E_e$ and radius $r$ respectively. 

The implicit second-order upwind scheme is employed for the discretization of the energy term, i.e.,
\begin{equation}
\frac{N_{i,j}^{l+1}-N_{i,j}^l}{\Delta t}=\frac{1}{2}\left[\frac{-b_{i,j+2}N_{i,j+2}^{l+1}+4b_{i,j+1}N_{i,j+1}^{l+1}-3b_{i,j}N_{i,j}^{l+1}}{2\Delta E}+ \frac{-b_{i,j+2}N_{i,j+2}^{l}+4b_{i,j+1}N_{i,j+1}^{l}-3b_{i,j}N_{i,j}^{l}}{2\Delta E}\right].
\end{equation}
where $i$ is the index of the spatial step and $j$ is the index of the energy step. After a few manipulation, we obtain
\begin{equation}
N_{i,j}^{l+1}=\left[\frac{-b_{i,j+2}N_{i,j+2}^{l+1}+4b_{i,j+1}N_{i,j+1}^{l+1}-3b_{i,j}N_{i,j}^{l+1}-b_{i,j+2}N_{i,j+2}^{l}+4b_{i,j+1}N_{i,j+1}^{l}}{4\Delta E}\Delta t+ N_{i,j}^l\right]\Big/\left[1+\frac{3b_{i,j}\Delta t}{4\Delta E} \right].
\end{equation}
Given the boundary condition $N_{i,j_{\rm max}}=0$ for any $l$, we can solve $N_{i,j}^{l+1}$ from $j=j_{\rm max}-1$ to $j=0$.

For the discretization of the spatial term, we adopt the finite volume method of the implicit scheme, which leads to
\begin{equation}
\frac{N_{i,j}^{l+1}-N_{i,j}^l}{\Delta t}=\frac{D_{i+1/2,j}r_{i+1/2}^2}{2\Delta r^2r_i^2}\left(N_{i+1,j}^l - N_{i,j}^l +N_{i+1,j}^{l+1} -N_{i,j}^{l+1} \right)-\frac{D_{i-1/2,j}r_{i-1/2}^2}{2\Delta r^2r_i^2}\left(N_{i,j}^l - N_{i-1,j}^l +N_{i,j}^{l+1} -N_{i-1,j}^{l+1} \right).
\end{equation}
where $r_{i+1/2}=r_i+\Delta r/2$, $r_{i-1/2}=r_i-\Delta r/2$, $D_{i+1/2,j}=(D_{i,j}+D_{i+1,j})/2$ and $D_{i-1/2,j}=(D_{i-1,j}+D_{i,j})/2$. This equation can be re-arranged into
\begin{equation}\label{eq:fvm}
\begin{split}
&-\frac{(D_{i,j}+D_{i-1,j})\Delta t}{4\Delta r^2}\left(1-\frac{\Delta r}{r_i}\right)N_{i-1,j}^{l+1}+\left[1+\frac{(D_{i,j}+D_{i+1,j})\Delta t}{4\Delta r^2}\left(1+\frac{\Delta r}{r_i}\right)+\frac{(D_{i,j}+D_{i-1,j})\Delta t}{4\Delta r^2}\left(1-\frac{\Delta r}{r_i}\right) \right]N_{i,j}^{l+1}\\
&-\frac{(D_{i,j}+D_{i+1,j})\Delta t}{4\Delta r^2}\left(1+\frac{\Delta r}{r_i}\right)N_{i+1,j}^{l+1}
=\frac{(D_{i,j}+D_{i-1,j})\Delta t}{4\Delta r^2}\left(1-\frac{\Delta r}{r_i}\right)N_{i-1,j}^{l}\\
&+\left[1-\frac{(D_{i,j}+D_{i+1,j})\Delta t}{4\Delta r^2}\left(1+\frac{\Delta r}{r_i}\right)-\frac{(D_{i,j}+D_{i-1,j})\Delta t}{4\Delta r^2}\left(1-\frac{\Delta r}{r_i}\right) \right]N_{i,j}^{l} + \frac{(D_{i,j}+D_{i+1,j})\Delta t}{4\Delta r^2}\left(1+\frac{\Delta r}{r_i}\right)N_{i+1,j}^{l}
\end{split}
\end{equation}
which is a tridiagonal system and can be easily solved. Note that the above equation is for a continuous $D(r)$. If  $D(r)$ follows the form of the step function as is shown in Eq.~(\ref{eq:D}), we have $D_{i-1/2,j}=D_{i,j}$ and $D_{i+1/2,j}=D_{i+1,j}$. Thus, all the terms of $(D_{i,j}+D_{i-1,j})$ in the numerator need to be be replaced by $2D_{i,j}$ and all the terms of $(D_{i,j}+D_{i+1,j})$ need to be replaced by $2D_{i+1,j}$ in Eq.~\ref{eq:fvm}. For the outer boundary where $i=i_{\rm max}$ (corresponding to a sufficient large distance to the pulsar), we impose $N_{i_{\rm max},j}=0$ for any $l$. For the inner boundary where $i=0$ or $r_1=0$, we utilize the symmetry of the system, i.e., $\partial N/\partial r=0$, leading to $N_{-1,j}=N_{1,j}$ and $D_{-1,j}=D_{1,j}$.

Particles are assumed to inject from $r=0$, which is embodied by the Dirac function $\delta (r)$ in Eq.~(\ref{eq:transport}). We adopt a rectangular function to approximate the Dirac function, i.e.,
\begin{equation}
Q(E_e,t)\delta(r)=\frac{Q(E_e,t)}{4\pi \Delta r^3}\times \left\{
\begin{array}{ll}
1, \quad {\rm i=0,1}\\
0, \quad i>1
\end{array}
\right.
\end{equation}
In our calculation, we set $\Delta t=5$yr, $\Delta r=0.25$pc with $r_{i_{\rm max}}=2.5\,$kpc. We divide energy in the logarithmic space with $\Delta \lg E=1/300$ between $E_0=0.1\rm GeV$ and $E_{j_{\rm  max}}=0.1\rm EeV$.  
  
\bibliography{ms_x.bib}

\end{document}